\documentstyle[preprint,aps]{revtex}
\title{Theory of Fermion to Boson Mappings: \\
 Old Wine in New Bottles}
\author{Joseph N.~Ginocchio and Calvin W.~Johnson}
\address{T-5, MS B283, Theoretical Division, Los Alamos National Laboratory,
Los Alamos, NM
87545}

\begin{document}

\maketitle
\begin{abstract}
\noindent
After a brief review of various mappings of fermion pairs to
bosons, we rigorously derive a general approach.  Following the
methods of Marumori and Otsuka, Arima, and Iachello, our approach begins
with mapping states and constructs boson representations that preserve
fermion matrix elements.  In several cases these representations factor into
finite, Hermitian boson images
times a projection or norm operator that embodies the
Pauli principle.  We pay particular attention to truncated boson spaces, and
describe general methods for constructing Hermitian and approximately finite
boson image Hamiltonians, including effective operator theory to account for
excluded states.  This method is akin to that of Otsuka, Arima, and Iachello
introduced in connection with the Interacting Boson Model,
but is more rigorous, general, and systematic.
\end{abstract}


\section{ Introduction}

The original attempt at bosonization of the nuclear many-fermion system was
motivated by collective particle-hole modes in nuclei\cite{BZ,Maru}.
Since that time the interacting boson model \cite{IBM} (IBM) has been
phenomenologically very successful in explaining low energy nuclear
spectroscopy for heavy nuclei.
The bosons in this model are thought to represent
monopole (J=0), quadrupole (J=2), and sometimes hexadecapole (J=4)
correlated pairs of valence nucleons in the shell model.
The IBM
Hamiltonian is Hermitian, usually has at most two-boson interactions,
and conserves boson number, reflecting the particle-particle, rather than
particle-hole, nature of the underlying fermion pairs.
While one can numerically
diagonalize the general IBM Hamiltonian, one of the strengths of IBM is
the existence of algebraic limits corresponding to the subgroups SU(3),
U(5), or O(6), with analytic expressions for excitation bands and
transition strengths, which encompass an enormous amount of
nuclear data.

The microscopic reasons for the success of such a simple model are elusive.
Otsuka, Arima, and Iachello, along with Talmi,
have used a mapping of the shell model Hamiltonian to the
IBM Hamiltonian \cite{OAIT,OAI} based on the
seniority model \cite{Talmi}, but these attempts have not done well for
well-deformed nuclei \cite{deformed}. For this reason we have revisited boson
mappings to see if we can understand the success of the IBM starting
from the shell model.

We will begin by sketching out various historic approaches to boson
map\-pings \cite{Review,RS}.  We then specificially follow Marumori \cite{Maru}
and Otsuka et al.\ \cite{OAIT,OAI} (OAI)
in our mapping procedure which maps fermion states into boson states and
construct boson operators that reproduce fermion matrix elements.
We give the boson representation of the
 Hamiltonian and review the result that, in the full Boson Fock space, it
factorizes into a boson image, which is the same as the Belyaev-Zelevinskii
Hamiltonian \cite {BZ}--in fact in the full space all mappings yield the same
results-- times a normalization operator which projects out the
spurious states.
However, since our goal is to understand the IBM, which only
deals with a few of the enormous degrees of freedom of the shell model,
we go on to discuss boson images in truncated spaces.  This, we shall see,
gives rigorous insight into the OAI mapping and shows how to  systematically
extend it.

\section{A brief history of boson mappings}

The fundamental goal is to solve the many-fermion Schr\"odinger
equation
\begin{equation}
\hat{H} \left | \Psi_\lambda \right \rangle = E_\lambda
 \left | \Psi_\lambda \right \rangle
\label{Schrodinger1}
\end{equation}
and find transition matrix elements between eigenstates,
$t_{\lambda \lambda^\prime} =
\left \langle \Psi_\lambda \left | \hat{T}
 \right | \Psi_{\lambda^\prime} \right \rangle$.  For fermion many-body
(shell-model)
basis states one often
uses Slater determinants, antisymmetrized products of single-fermion
wavefunctions which we can write using Fock creation operators:
$a^{\dagger}_{i_{1}} \cdot \cdot \cdot
a^{\dagger}_{i_{n}} \left | 0 \right \rangle $ for $n$ fermions.
For an even number of fermions one can instead construct states from products
of fermion pairs,
\begin{equation}
\left | \Psi_\beta \right \rangle
= \prod_{m=1}^N\hat{A}^\dagger_{\beta_m} \left | 0 \right \rangle;
\label{StateDefn}
\end{equation}
if the number of fermion is fixed at $n$ then $m$ runs from 1 to $N=n/2$.  As
the fermion Fock
space may be so large as to make direct solution intractable, the idea of
a boson mapping is to replace the fermion operators with boson operators,
{\em using only a minimal number of boson degrees of freedom}, that approximate
the spectrum and transition matrix elements of the original fermion problem.
There are two approaches to boson mappings which we now review.

The first approach, epitomized
in nuclear physics by Belyaev and Zelevinskii (BZ) \cite{BZ},
is to map fermion operators to boson operators
so as to preserve the original algebra. Specifically, consider a
space with $2\Omega$ single-fermion states; $a^\dagger_i, a_j$ signify
fermion creation and annihilation operators.
The set of all bilinear fermion operators,
$a_ia_j, a^\dagger_k a^\dagger_l, a^\dagger_i a_j$, form the Lie algebra of
${\rm SO}(4\Omega)$, as embodied by the commutation
relations
\begin{eqnarray}
\left [ a_i a_j, a_k a_l \right ] = 0  \\
\left [ a_i a_j, a^\dagger_k a^\dagger_l \right ] =
\delta_{il} \delta_{jk} +
\delta_{ik} a^\dagger_l a_j
+\delta_{jl} a^\dagger_k a_i
- (i \leftrightarrow j)
 \\
\left [ a_i a_j, a_k^\dagger a_l \right ] =
\delta_{jk} a_i a_l - (i \leftrightarrow j)
 \\
\left [ a^\dagger_i a_j, a^\dagger_k a_l \right ] =
\delta_{jk} a^\dagger_i a_l -\delta_{il} a^\dagger_k a_j
\end{eqnarray}
At this point it is convenient to introduce collective fermion
pair operators
\begin{equation}
\hat{A}^\dagger_\beta \equiv  { 1 \over \sqrt{2}} \sum_{ij}
\left ( {\bf A}^\dagger_\beta \right )_{ij}
{a}^\dagger_i {a}^\dagger_j.
\end{equation}
We always choose the  $\Omega(2\Omega-1)$ matrices ${\bf A}_\beta$ to be
antisymmetric so as to
preserve the underlying fermion statistics, thus eliminating the need
later on to distinguish between `ideal' and `physical' bosons.
We also assume the following normalization and completeness relations for
the matrices:
\begin{eqnarray}
{\rm tr\,} {\bf A}_\alpha {\bf A}^\dagger_\beta = \delta_{\alpha \beta}; \\
\sum_\alpha \left ({\bf A}^\dagger_\alpha\right )_{ij}
\left ({\bf A}_\alpha\right )_{j^\prime i^\prime}
={1\over 2} ( \delta_{i i^\prime} \delta_{j j^\prime}
-\delta_{i j^\prime}  \delta_{j i^\prime} )  .
\label{completeness}
\end{eqnarray}
Generic one- and two-body
fermion operators we represent by
$ \hat{T} \equiv
\sum_{ij} T_{ij} {a}^\dagger_i {a}_j$,
$\hat{V} \equiv \sum_{\mu \nu} \left \langle \mu
\right | V \left |  \nu \right \rangle
\hat{A}^\dagger_\mu \hat{A}_\nu,$
where $T_{ij} = \left \langle i \right | \hat{T} \left | j \right \rangle$;
from such operators one can construct a fermion Hamiltonian $\hat{H}$.
Now one has the following commutation relations:
\begin{eqnarray}
\left [ \hat{A}_\alpha , \hat{A}_\beta \right ] =
\left [ \hat{A}^\dagger_\alpha , \hat{A}^\dagger_\beta \right ] = 0;
\label{commute1} \\
\left [ \hat{A}_\alpha , \hat{A}^\dagger_\beta \right ] =
\delta_{\alpha \beta} -2 \sum_{ij}
\left ( {\bf A}^\dagger_\beta {\bf A}_\alpha \right )_{ij}
a^\dagger_i a_j ; \\
\left [ \hat{A}_\alpha , \hat{T} \right ] =
2\sum_\beta {\rm tr \,} \left (
{\bf A}_\alpha {\bf T} {\bf A}^\dagger_\beta \right ) \hat{A}_\beta
\\
\left [ \hat{T}_1, \hat{T}_2 \right ]
= \sum_{ij} \left ( {\bf T}_1{\bf T}_2 -{\bf T}_2{\bf T}_1 \right)_{ij}
a^\dagger_i a_j
\label{commutelast}
\end{eqnarray}
The method of Belyaev and Zelevinskii is to find boson images of the
bifermion operators,
\begin{eqnarray}
(\hat{A}^\dagger_\mu)_B = b^\dagger_\mu + \sum_{\alpha \beta \gamma}
x_\mu^{\alpha \beta \gamma} b^\dagger_\alpha b^\dagger_\beta b_\gamma
+ \sum_{ \alpha \beta \gamma \delta \epsilon }
x_\mu^{\alpha \beta \gamma \delta \epsilon }
b^\dagger_\alpha b^\dagger_\beta  b^\dagger_\delta b_\gamma b_\epsilon
+ \ldots
\label{BZpair}
\\
(\hat{A}_\mu)_B = (A^\dagger_\mu)_B^\dagger \\
(\hat{T})_B =
\sum_{\alpha \beta} y^{\alpha \beta} b^\dagger_\alpha b_\beta
+\sum_{\alpha \beta \gamma \delta } y^{\alpha \beta \gamma \delta }
 b^\dagger_\alpha b^\dagger_\gamma b_\beta b_\delta
+ \ldots
\end{eqnarray}
where $b_\alpha, b^\dagger_\beta$ are boson creation and annihilation
operators, $[ b_\alpha, b^\dagger_\beta ] = \delta_{\alpha \beta}$,
with the coefficients $x, y$
chosen so that the images $(A^\dagger_\mu)_B, (A_\nu)_B, (T)_B$
 have the same commutation relations as in
(\ref{commute1})-(\ref{commutelast}).
Because the algebra is exactly matched, if one builds boson states in
exact analogy to the fermion states
then the full boson Fock space is not spanned and one does not have
nonphysical or spurious states.

In the full boson Fock space, that is,
no truncation of the boson degrees of freedom, the image
of one body operators is finite and given quite simply by
$(\hat{T})_B  = 2 \sum_{\alpha \beta} {\rm tr \,} ({\bf A}_\alpha {\bf T}
{\bf A}^\dagger_\beta ) b^\dagger_\alpha b_\beta$.
Since in the full space any fermion Hamiltonian can be written in terms of
one-body operators,
the boson image of a finite fermion Hamiltonian will be finite in the full
space.  The states that one must use then are built from the boson
representations of the fermion pairs given in
(\ref{BZpair}) which will not just be products of bosons but will include
exchange terms.  For example, for two bosons and using (\ref{BZpair}),
\begin{equation}
\hat{A}^\dagger_\alpha \hat{A}^\dagger_\beta
\left | 0 \right \rangle \rightarrow
\left ( b^\dagger_\alpha b^\dagger_\beta
+ x^{\sigma \tau\beta}_{\alpha}  b^\dagger_\sigma b^\dagger_\tau \right )
\left | 0 \right ).
\end{equation}
These exchange terms are due to the antisymmetry.  We shall take care of
such exchange effects by introducing a norm operator in the boson space.

For {\it truncated spaces}, however, blind application of the boson
representations of the pair operators
(\ref{BZpair}) will produce states outside the truncated space.
Therefore, if these
states are omitted, then the Hamiltonian needs to be renormalized to account
for
these omissions, which in general leads to an infinite BZ Hamiltonian.
Marshalek \cite{Marshalek} points out that there exist mappings that
are both finite and Hermitian,
but these in general require projection operators to eliminate
spurious states.  We will regain this result later on in this paper.

A variant of Belyaev-Zelevinskii that also preserves commutation relations is
the Dyson mapping \cite{Dyson}:
\begin{eqnarray}
\hat{A}_\alpha \rightarrow b_\alpha; \\
\hat{A}^\dagger_\beta \rightarrow b^\dagger_\beta
- 2\sum_{\lambda \mu \nu}
{\rm tr \,} ( {\bf A}_\lambda {\bf A}^\dagger_\beta
{\bf A}_\mu {\bf A}^\dagger_\nu )
b^\dagger_\lambda b^\dagger_\mu b_\nu\\
\hat{T} \rightarrow 2 {\rm tr \,}
\sum_{\alpha \beta} ({\bf A}_\alpha {\bf T}
{\bf A}^\dagger_\beta ) b^\dagger_\alpha b_\beta.
\end{eqnarray}
The operators are then clearly finite; on the other hand they are just as
clearly non-Hermitian.  From a computational viewpoint non-Hermiticity is
only a minor
barrier, but it is an obstacle to an understanding of the microscopic
origin of Hermitian IBM Hamiltonians.  Furthermore, unless the truncated pairs
constitute a subalgebra, under truncation, renormalization will produce an
infinite expansion, although it may be possible that this expansion may be more
convergent than the BZ expansion.

The second major approach, pioneered by Marumori \cite{Maru}, is to map fermion
states and construct boson representation operators that preserve matrix
elements.  The original work of Marumori, however, focused on particle-hole
excitations
and so the number of pairs and consequently bosons were not fixed. However,
this
method can be applied to particle-particle pairs as well.

Marumori constructs the norm matrix
\begin{equation}
{\cal N}_{\alpha \beta} = \left \langle \Psi_\alpha | \Psi_\beta
\right \rangle
\end{equation}
and then the Usui operator
\begin{equation}
 U = \sum_{\alpha, \beta; n}
\left | \Phi_\beta \right )
\left ( {\cal N}\right)^{-1/2}_{\beta \alpha}
\left \langle \Psi_\alpha \right |
\end{equation}
where
bosons states are constructed in strict
analogy to the fermion states,
\begin{equation}
\left | \Phi \right \rangle =
\prod_{m=1}^N b^{\dagger}_{\beta_m} \left | 0 \right \rangle.
\end{equation}
Then the Marumori expansion of any fermion operator is
\begin{equation}
O_B =  U O_F  U^\dagger.
\end{equation}
We shall show that, in the full space, these boson representation operators
factor into a finite boson image times a norm operator.  Furthermore, the boson
image of the Hamiltonian is the same as the BZ image, and hence, the two
methods are equivalent in the full space.

Otsuka, Arima, and Iachello (OAI), along with Talmi \cite{OAIT,OAI},
investigated the microscopic origins of the Interacting Boson Model
through boson mappings.  Although they also mapped states, they differed
from Marumori in some key details.  First of all, they built states built
on a fixed number of particle-particle, not particle-hole, pairs.  In addition,
the space was truncated to include only one monopole ($J^\pi = 0^+$) and
quadrupole ($J^\pi = 2^+$) pair.
These states were orthogonalized based on seniority.
That is, they constructed, for $2N$ fermions,
 low-seniority basis states of $S$ and $D$ fermion pairs,
$\left | S^{N-n_d} D^{n_d} \right \rangle$, and then orthonormalized the
states such that the zero-seniority state is mapped to itself, and
states of higher seniority $v$ were orthogonalized against states of lower
seniority,
\begin{equation}
\left | v\right ) \rightarrow \left | ``v" \right ) =\left | v \right )
+ \left | v-2 \right ) + \left | v -4 \right ) + \ldots
\end{equation}
Then OAI calculated the matrix elements
$ \left \langle `` S^{N-n_d^\prime} D^{n_d^\prime } "\left |
H_F \right | `` S^{N-n_d} D^{n_d} "\right \rangle$ for
$n_d, n_d^\prime = 0,1,2$ and obtained the coefficients
for their one plus two-boson Hamiltonian. These coefficients have an implicit
$N$-dependence
(and for large $N$ and arbitrary systems such matrix elements
are not trivial to calculate, especially in analytic form!)
and thus a many-body dependence. At first sight this is not entirely
unreasonable as it is well known the
IBM parameters change substantially as a function of the number of
bosons, even within a major `shell'.
Nonetheless the OAI mapping has three drawbacks. The first is that
it is not clear how to systematically calculate many-body contributions
beyond that contained in the OAI prescription, whereas the  method we shall
describe is fully and rigorously systematic.
The second is that the OAI
prescription can induce many-body effects where none are needed.
This point will be illustrated in
section \ref{ApproxImages}.  Thirdly, only the $n_d = 0,1,2$
space is exactly mapped, but very deformed systems will involve large $n_d$.
In fact, for an axial rotor limit, the average number of d-bosons in the ground
 state band is 2/3 the total number of bosons.  Correcting this by also mapping
matrix elements with $n_d > 2$ will involve many-body terms.

As an alternative to OAI, Skouras, van Isacker, and Nagarajan
\cite{democratic} proposed a
``democratic'' mapping where the orthogonalization is based on
eigenvectors of the norm matrix rather than seniority.

In what follows we present a unified state-mapping
method and obtain four strong results.  First we derive matrix elements of the
fermion operators in the pair basis (2).  Second, we give
general expressions for fermion matrix elements via boson representations.
Third, we show how in several cases one can have exact, finite, and
Hermitian boson images of fermion operators.   Finally, we show how to
extend  the OAI and democratic mappings in a systematic and
rigorous fashion, and illustrate how the choice of orthogonalization
can affect the many-body dependence of the boson images.

\section{Matrix elements of fermion-pair states}

The starting point of any state-mapping method is the calculation of
matrix elements of fermion operators between states constructed from
fermion pairs of the form (\ref{StateDefn}), including the overlap:
$\left \langle \Psi_\alpha | \Psi_\beta \right \rangle $,
$\left \langle \Psi_\alpha \left | \hat{H} \right |
 \Psi_\beta \right \rangle $,
$\left \langle \Psi_\alpha \left | \hat{T} \right |
 \Psi_\beta \right \rangle $, and so on.
These matrix elements are much more difficult to compute
 than the corresponding matrix
elements between Slater determinants.  As we shall show, however, full and
careful attention paid to the problem of calculation matrix elements can
yield powerful results.
Silvestre-Brac and Piepenbring \cite{Silvestre},
laboriously using commutation relations, derived a  Wick theorem for
fermion pairs.   Rowe, Song and Chen \cite{Rowe}
using `vector coherent states'
(we would say fermion-pair coherent states) found matrix elements between
pair-condensate wavefunctions, states of the form
$\left ( \hat{A}^\dagger \right )^N \left | 0 \right \rangle.$
Using a theorem by Lang et al.\ \cite{Lang}, we have
generalized the method of Rowe, Song and Chen and
recovered (actually
discovered independently) the
expressions of Silvestre-Brac and Piepenbring.

To derive the generalized Wick's theorem, we use
the following theorem (its
proof, which requires fermion coherent states and integration over
Grassmann variables, is found in Appendix A of \cite{Lang}):

Let $\hat{U}$ be an operator of the form
\begin{equation}
\hat{U} \equiv
\exp \left ( \hat{h}(n) \right ) \cdots
\exp \left ( \hat{h}(2) \right )
\exp \left ( \hat{h}(1) \right ),
\label{U-def}
\end{equation}
where each $\hat{h}(t)$ is of the form
\begin{equation}
\hat{h}(t) = \sum_{ij} \left[ T(t)_{ij} \hat{a}^\dagger_i \hat{a}_j
+ B(t)_{ij} \hat{a}_i \hat{a}_j  +
A^*(t)_{ji} \hat{a}^\dagger_i \hat{a}^\dagger_j \right].
\end{equation}
Again, the matrices ${\bf A}$, ${\bf B}$ are antisymmetric (${\bf T}$
is not, in general), and are of dimension $2 \Omega \times 2 \Omega$.
We introduce the $4 \Omega \times 4 \Omega$ matrix representation of $\hat{U}$
in the basis of the single-particle Fock operators $a^{\dagger}_i , a_j$.
Then:
\begin{equation}
\left (
\matrix{
{\bf U}_{11} &  {\bf U}_{12} \cr
{\bf U}_{21} & {\bf U}_{22} \cr}
\right ) = \prod_{t=1}^n
\exp \left (
\matrix{
{\bf T}(t) & 2 {\bf A}^\dagger(t) \cr
2\bf{ B}(t) & -{\bf T}^T(t) \cr}
 \right )  .
\end{equation}
The conclusion of the theorem is the vacuum expectation value of $\hat{U}$:
\begin{equation}
\left \langle 0 \right | \hat{U} \left | 0 \right \rangle
= \sqrt{ \det \left ( {\bf U}_{22} \right ) }
\exp \left ( {1\over 2} \sum_{t} {\rm tr} {\bf T}(t) \right ).
\label{VEV}
\end{equation}

Appropriate derivatives of (\ref{VEV})   bring down
pair creation and annihilation operators, leading to the
desired matrix elements.  We  do this in some detail for the
overlap matrix elements
$\left \langle \Psi_\alpha \right | \left . \Psi_\beta \right \rangle$ for the
states given in (2);
then we  simply give the results for matrix elements of one- and
two-body operators which are found in the same way.
If one begins with
\begin{equation}
\hat{U} =
\prod_{i=1}^N  \exp \left(  \epsilon_{ \alpha_i}\hat{A}_{\alpha_i} \right )
\prod_{i=j}^N \exp \left(  \epsilon_{ \beta_j}\hat{A}_{\beta_j}^\dagger\right),
\end{equation}
where $\hat{A}_{\alpha} , \hat{A}^{\dagger}_{\beta}$ are defined in (7), then
the overlap is
\begin{equation}
\left \langle \Psi_\alpha \right | \left . \Psi_\beta \right \rangle
= \left .
\left (
\prod_{i = 1}^N
{\partial^2 \over
{
\partial \epsilon_{\alpha_i} \,
\partial \epsilon_{\beta_i}
}
} \right )
\left \langle 0 \right | \hat{U} \left | 0 \right \rangle
\right |_{ {\rm all} \, \epsilon {\rm 's} = 0}
\label{diffU}
\end{equation}

In applying the theorem the matrix
algebra is straightforward; keeping only
terms linear in $\epsilon_{\alpha_i}$ etc.~one finds
that for the overlap $T(t) = 0$, and
\begin{equation}
{\bf U}_{22} = {\bf 1} +
2 \sum_{ij}
\epsilon_{\alpha_i} \epsilon_{\beta_j}
{\bf A}_{\alpha_i} {\bf A}^{ \dagger}_{\beta_j}.
\end{equation}
Next, insert ${\bf U}_{22}$ into (\ref{VEV}), use
$\det {\bf U}_{22} = \exp {\rm tr} \ln {\bf U}_{22}$ and
expand the logarithm to arrive at the generating function
\begin{equation}
\left \langle 0 \right | \hat{U} \left | 0 \right \rangle
=
\exp \left (
 \sum_{n=1}^\infty { (-2)^{n-1} \over n}
{\rm tr} \left (
 \sum_{ij}
\epsilon_{\alpha_i} \epsilon_{\beta_j}
{\bf A}_{\alpha_i} {\bf A}^{ \dagger}_{\beta_j} \right )^n
\right ).
\label{VEV_trace}
\end{equation}
Finally, one applies (\ref{diffU}).

The result is constructed from
objects of the form
$ {\rm tr} \left (
{\bf A}_{\alpha_1} {\bf A}^{ \dagger}_{\beta_1}
{\bf A}_{\alpha_2} {\bf A}^{ \dagger}_{\beta_2}
\ldots
{\bf A}_{\alpha_k} {\bf A}^{ \dagger}_{\beta_k}\right )$,
which we term a {\it k-contraction} (related to the contractions
of $k$ phonons in \cite{Silvestre})
and represent with the notation
$c_k[ \alpha_1, \beta_1; \alpha_2, \beta_2;\ldots;
\alpha_k, \beta_k]$ (from the order it is clear that the $\alpha$'s
represent final states and the $\beta$'s initial states).
Then
$\left \langle \Psi_\alpha \right | \left . \Psi_\beta \right \rangle$ is
a finite sum of products of $k$-contractions; each term in the sum is
a product of $m_1$ 1-contractions, $m_2$ 2-contractions, \dots,
and $m_N$ $N$-contractions, with the $\left \{ m_i \right \}$ taking all
possible values under the
constraint that $\sum_k k  m_k = N$.  For example, there will be
terms consisting of $N$ 1-contractions, terms consisting of $N-2$
1-contractions and one 2-contraction, and
so on, up to terms consisting of 1 $N$-contraction.
The full expression is
\begin{equation}
\left \langle \Psi_\alpha \right | \left . \Psi_\beta \right \rangle =
\sum_{\cal P} \sum_{\ell = 0}^N
w_\ell^0 \left( \alpha_1,\alpha_2, \ldots, \alpha_\ell;
\beta_1, \ldots, \beta_\ell \right ) \left ( (N-\ell)! \right)^{-1}
\prod_{k = \ell+1}^N c_1 [ \alpha_k, \beta_k ]
\label{norm}
\end{equation}
where ${\cal P}$ means the sum is over all permutations of
$\left \{ \alpha_i \right \}$ and
$\left\{ \beta_j \right \}$.
The coefficients $w^0_\ell$ are given by
\begin{eqnarray}
w^0_l
\equiv
\sum_{m_2,\ldots,m_\ell}
z_\ell(m_2,\ldots,m_l)
\nonumber \\
\times
\underbrace{
c_2[\alpha_1, \beta_1; \alpha_2, \beta_2] \,
c_2[\alpha_3, \beta_3; \alpha_4, \beta_4] \,
\ldots
c_2[\alpha_{m_2 -1}, \beta_{m_2-1}; \alpha_{m_2}, \beta_{m_2}] \,
}_{m_2}
 \\
\times
\underbrace{
c_3[\alpha'_1, \beta'_1; \alpha'_2, \beta'_2;
\alpha'_3, \beta'_3]
 \ldots
c_3[\alpha'_{m_3-2}, \beta'_{m_3-2}; \alpha'_{m_3-1}, \beta'_{m_3-1};
\alpha'_{m_3}, \beta'_{m_3}]
}_{m_3} \times \ldots \nonumber
\end{eqnarray}
The organization of  (\ref{norm}) is such as to ease the
interpretation in bosons in the next section.

The coefficients $z_\ell(m_2,\ldots,m_\ell)$ are found through the
expansion of (\ref{VEV_trace}):
\begin{equation}
z_\ell(m_2,\ldots,m_\ell) =
 (-2)^{\ell-M} \,
\prod_{k > 1}
 { 1 \over { k^{m_k} m_k! } },
\end{equation}
where
$\ell = \sum_{k>1} k m_k$ and $M = \sum_{k>1} m_k$, and $w^0_0=z_0=1$
(hence the normalization chosen above). Note that we have set
$m_1 = N-\ell$ and that the $w_\ell^0$ have no explicit dependent on  $N$.

To illustrate, consider a single $j$-shell fermion space, with
$\Omega = j + {1 \over 2}$, and further consider only the fermion pair
with total $J= 0$:
$A_0^\dagger = (2)^{- 1/2} \left [ a^\dagger_j \otimes a^\dagger_j \right ]_0$;
the matrix ${\bf A}^\dagger_0$ is antidiagonal,
$\left ({\bf A}^\dagger_0\right)_{m m^\prime} = (2\Omega)^{-1/2}
(-1)^{j - m} \delta_{m, -m^\prime}$.  Then the $k$-contraction
$c_k  = {\rm tr\,} \left ( {\bf A}_0{\bf A}_0^\dagger
\right )^k = \left ( 2 \Omega \right )^{-k+1}$.
Following the above definitions,
$$
z_2(1) = -1, \, z_3(0,1) = 4/3, \, z_4(2,0,0) = 1/2, \,
$$
$$
z_4(0,0,1) = -2,  \, z_5(1,1,0,0) = -4/3, \, z_5(0,0,0,1) = 16/5, \,{\rm etc.}
$$
and
\begin{equation}
w^0_2 = -{1 \over 2\Omega}, \, w^0_3 = {1 \over 3\Omega^2}, \,
w^0_4 = {1 \over 8\Omega^2 } - {1 \over 4\Omega^3}, \,
w^0_5 = -{1 \over 6\Omega^3} + {1 \over 5\Omega^4} .
\label{w_coef}
\end{equation}
Then from equation (\ref{norm}), including the sum over all permuations
which gives a factor $N!^2$, one obtains for example the norm
of the state with 5 $J=0$ pairs:
$$
\left \langle 0 \left |
A_0^5 (A_0^\dagger)^5 \right | 0 \right \rangle
= 5! \left ( 1 - {10 \over \Omega} +{ 35 \over \Omega^2}
- {50 \over \Omega^3}  + {24 \over \Omega^4} \right ),
$$
which agrees with the general result from the commutation relation
$\left [ A_0, A^\dagger_0 \right ] = 1 - \hat{N}/ \Omega $ ($\hat{N}$ is the
fermion number operator), that is, the norm is
$N! \Omega! / (\Omega - N)! \Omega^{N-1}$.

In the same way as for the overlap
one can derive the matrix elements for one- and two-body
operators $\hat{\cal O}^{1,2}$ (and for general $n$-body operator, if so
desired):
\begin{equation}
\left \langle \Psi_\alpha \right | \hat{\cal O}^{1,2} \left | \Psi_\beta \right
\rangle =
\sum_{\cal P} \sum_{\ell = 1}^N \left ( (N-\ell)! \right)^{-1}
\sum_{k = 1}^\ell  \tilde{w}_k^{1,2} [ {\cal O}^{1,2};
\alpha_1, \ldots, \beta_k ]
w_{\ell -k}^0
\prod_{m = \ell+1}^N c_k [ \alpha_k, \beta_k ] .
\end{equation}
For a one-body operator $\hat{T}$ we define ${\bf A}_\alpha {\bf T}
= {\bf A}_{\alpha(T)}$; then
\begin{equation}
\tilde{w}^1_k (T; \alpha_1, \alpha_2, \ldots, \beta_k) \equiv
 2  (-2)^{k-1} c_k [
\alpha_1(T), \beta_1;
\alpha_2, \beta_2;
\ldots;\alpha_k, \beta_k ]   .
\end{equation}
For a two-body operator $\hat{V}$
\begin{eqnarray}
\tilde{w}^2_k (V; \alpha_1, \alpha_2, \ldots, \beta_k) \equiv \nonumber \\
\sum_{ \mu \nu }
\left \langle \mu \right | \hat{V} \left | \nu \right \rangle  (-2)^{k-1}
\sum_{l = 1}^k
\left (
c_l[ \mu,\beta_1; \alpha_2, \beta_2; \ldots; \alpha_l, \beta_l]
c_{k -l +1} [ \alpha_1,\nu; \alpha_{l+1}, \beta_{l+1} ;
\ldots; \alpha_k, \beta_k ] \right . \nonumber \\
\left . -2 (\delta_{l,1} - 1)
c_{k+1} [
\mu, \beta_1; \alpha_2, \beta_2; \ldots;
\alpha_l, \beta_l; \alpha_1,\nu;
\alpha_{l+1}, \beta_{l+1}; \ldots; \alpha_k, \beta_k ]
\right ).
\end{eqnarray}
Returning to our illustration, first
consider the the number operator $\hat{N}$, which is represented
by just the unit matrix; in this case, again restricting ourselves to
$J=0$ pairs,  $\tilde{w}^1_k = 2/(-\Omega)^{k-1}$.
Similarly for the pairing interaction $V = A_0^\dagger A_0$,
$$
\tilde{w}^2_k = \left ( 1 \over -\Omega \right )^{k -1}
\left ( k \left( 1 - {1\over \Omega} \right ) + {1 \over \Omega} \right ).
$$
The reader is invited to check that these coefficients reproduce the correct
matrix elements.

Therefore, given any two states constructed from fermion pairs and the
matrices representing those constituent pairs, the above formulas give
exactly the overlap and the matrix element for one- and two-body operators.
For a pair-condensate wavefunction, the matrix elements can be found
quickly through recursion \cite{Silvestre,Rowe}.

Throughout this paper we will discuss the effect of truncating the
boson Fock space, that is taking a restricted number of boson species from
which to construct states and operators, on boson mappings.
But which boson species should we keep?  In regards to this question we
 merely wish to comment that Rowe, Song and Chen give a
variational principle \cite{Rowe} which seems useful in this regard, and
is probably related in some approximation to the Hartree-Fock-Bogoliubov
states that Otsuka and Yoshinaga \cite{OY} use in their mapping of
deformed nuclei.

\section{Boson representations of fermion matrix elements}

We now want to translate the fermion matrix elements into boson space.
We take the simple mapping of fermion states into boson
states
\begin{equation}
 \left | \Psi_\beta \right \rangle
\rightarrow  \left | \Phi_\beta \right )
= \prod_{m=1}^N {b}^\dagger_{\beta_m} \left | 0 \right ),
\end{equation}
where the ${b}^\dagger$ are boson creation operators.
We construct boson operators that preserve matrix elements, introducing boson
operators
$\hat{\cal T}_B$, $\hat{\cal V}_B$, and most
importantly the {\it  norm operator} $\hat{\cal N}_B$
 such that
$ \left ( \Phi_\alpha \right | \hat{\cal T}_B \left | \Phi_\beta \right )
= \left \langle \Psi_\alpha \right| \hat{T} \left | \Psi_\beta \right \rangle$,
$
\left ( \Phi_\alpha \right | \hat{\cal V}_B \left | \Phi_\beta \right )
= \left \langle \Psi_\alpha \right| \hat{V} \left | \Psi_\beta \right \rangle.
$
and
$ \left ( \Phi_\alpha \right | \hat{\cal N}_B \left | \Phi_\beta \right )
= \left \langle \Psi_\alpha \right. \left | \Psi_\beta \right \rangle$.
We term $\hat{\cal T}_B$,$\hat{\cal V}_B$ the boson {\it representations} of
the fermion operators $\hat{T}$, $\hat{V}$.  The boson basis
$\left | \Phi_{\beta} \right )$ is an orthogonal basis.
The fermion norm operator in the boson space will be given by
\begin{equation}
\hat{\cal{N}}_B = \sum \; \frac{\langle \Psi_{\alpha} | \Psi _{\beta}
\rangle}{(N!)^2} \; b^{\dagger}_{\alpha} ... b^{\dagger}_{\alpha_{N}}
b_{\beta_{1}} ... b_{\beta_{N}}
\end{equation}
where the coefficients are given by (\ref{norm}).  Because the matrices
${\bf A}_\alpha, {\bf A}^\dagger_\beta$ are orthogonal (see (8)),
the one-contraction is simply $c_1 \left[ \alpha_k , \beta_k  \right] =
\delta_{\alpha_{k}, \beta_{k}}$.  Using the fact that $b^{\dagger}_{\alpha}
b_{\alpha} = \hat{N}$, the number operator, we find the `linked-cluster'
(\`a la Kishimoto and Tamura\cite{KT,Sakamoto}
although with differences) expansion of the
representations to be of the form
\begin{equation}
\hat{\cal N}_B = 1 + \sum_{\ell=2}^\infty
\sum_{\left \{ \sigma, \tau \right \} }
w_\ell^0( \sigma_1, \ldots, \sigma_l; \tau_1, \ldots, \tau_l)
\prod_{i=1}^\ell  b^\dagger_{\sigma_i}
\prod_{j=1}^\ell  b_{\tau_j} .
\label{NormExpand}
\end{equation}
and similarly for $\hat{\cal V}_B, \hat{\cal T}_B$.  In the norm operator
the $\ell$-body terms express the
fact that the fermion-pair operators do not have exactly bosonic
commutation relations, and act to enforce the Pauli principle.

In the example of a single $j$-shell given in the previous section,
using the coefficients (\ref{w_coef}) and with the mapping
$A_0^\dagger \rightarrow s^\dagger$, the purely $s$-boson part of the
norm is
\begin{equation}
\hat{\cal N}_B = 1  - {1 \over 2\Omega} s^\dagger s^\dagger s s
+ {1 \over 3\Omega^2}   s^\dagger  s^\dagger  s^\dagger  sss
+ \left (  {1 \over 8\Omega^2 } - {1 \over 4 \Omega^3}  \right )
s^\dagger  s^\dagger  s^\dagger  s^\dagger ssss
+ \ldots
\end{equation}
which again yields the correct matrix elements.

The norm operator can be conveniently and compactly
expressed in terms of bosons by using the fermion norm matrix
(\ref{diffU}) in terms of
the fermion generating function (\ref{VEV_trace}).
Taking derivatives with respect to the
$\epsilon_{\beta_{i}}$ in (\ref{diffU})
is like contracting $b^{\dagger}_{\beta_{i}}$
with a $b_{\beta_{i}}$, and with respect to $\epsilon_{\alpha_{i}}$ is like
contracting a $b_{\alpha_{i}}$ with a $b^{\dagger}_{\alpha_{i}}$.  Hence the
norm matrix is just the generating function (\ref{VEV_trace})
with $\epsilon_{\alpha_{i}} \rightarrow b^{\dagger}_{\alpha_{i}}$ and
$\epsilon_{\beta_{j}} \rightarrow b_{\beta_{j}}$.
However, since these bosons do not
commute with one another, we must take the normal order:
\begin{equation}
\hat{\cal N}_B
= \, \colon \exp \left ( - { 1\over 2}
\sum_{k=2}^\infty
{ (-1)^{k} \over k}\hat{C}_k
\ \right ) \colon
\label{CompactNorm}
\end{equation}
where the colons `:' refer to normal-ordering of the boson operators, and
$\hat{C}_k = 2 \colon {\rm tr \,} {\bf P}^k \colon$ is the
$k$th-order Casimir of ${\rm SU}(2\Omega)$, with
${\bf P} = \sum b^\dagger_\sigma b_\tau {\bf A}_\sigma {\bf A}^\dagger_\tau$
(the trace is over the matrices and not the boson Fock space).
This norm operator,
which takes into account the exchange terms in the BZ expansion
of a fermion pair given in (\ref{BZpair}), is found in Ref.~\cite{Doba}.

Similarly --- and this is a new result we have not seen elsewhere in
the literature  --- the
representations $\hat{\cal T}_B, \hat{\cal V}_B$ can also be written
in compact form:
\begin{eqnarray}
\hat{\cal T \rm}_B & = & 2 \sum_{\sigma,\tau} \colon {\rm tr \,
}
\left[ {\bf A}_\sigma {\bf T} {\bf A}^\dagger_\tau {\bf G} \right]
b^\dagger_\sigma b_\tau \hat{ \cal N}_B \colon
\label{OneBodyImage}
\\
\hat{\cal V}_B & = &
\sum_{\mu, \nu}
\left \langle \mu \left | V \right | \nu \right \rangle
\sum_{\sigma, \tau} \colon \left \{
{\rm tr \,} \left [ {\bf A}_\sigma {\bf A}^\dagger_\mu {\bf G} \right ]
{\rm tr \,} \left [ {\bf A}_\nu {\bf A}^\dagger_\tau {\bf G} \right ]
\right. \nonumber \\
\hspace*{1.00in} & + & \left.4 \,{\rm tr \,} \left [
{\bf A}_\sigma {\bf A}^\dagger_\mu {\bf P G A}_\mu {\bf A}^\dagger_\tau
{\bf G} \right ] \right \}
b^\dagger_\sigma b_\tau  \hat{ \cal N}_B \colon ,
\label{TwoBodyImage}
\end{eqnarray}
where ${\bf G} = ({\bf 1} + 2{ \bf P})^{-1}$.
These compact forms are useful for formal manipulation.  Furthermore
they have  the powerful property of exactly expressing the
fermion matrix elements under {\it any} truncation, a fact not previously
appreciated in the literature even for the norm operator \cite{Doba}.
By this we mean the following: suppose we truncate our fermion Fock space
to states constructed from a restricted set of pairs
$\left \{ \bar{\sigma } \right \}$. Such a truncation need {\it not} correspond
to any subalgebra.   Then the representations in the corresponding
truncated boson space, which still exactly reproduce the fermion matrix
elements and which we denote by $\left [ {\cal N }_B \right ]_T$ etc.,
are  the same as those given above, retaining only the `allowed'
bosons with unrenormalized coefficients.  For example
\begin{equation}
\left[{\cal N \rm}_B \right]_T = \colon \exp \left(-\frac{1}{2}
\sum^{\infty}_{k=2} \frac{(-1)^k}{k} \left[\hat{C}_k\right]_T\right):
\end{equation}
where
\begin{equation}
[\hat{C}_k]_T = 2  \colon {\rm tr \, } (\left [{\bf P}\right ]_T )^k
 \colon \, ,\, \left [{\bf P}\right ]_T =  \sum_{\bar{\sigma}\bar{\tau}}
b^{\dagger}_{\bar{\sigma}} b_{\bar{\tau}} {\bf
A}_{\bar{\sigma}} {\bf A}^{\dagger}_{\bar{\tau}}.
\end{equation}
This invariance of the coefficients under truncation will
not hold true for the boson {\it images} introduced below.

With the boson representations of fermion operators in hand, one can
express the fermion Schr\"odinger equation (\ref{Schrodinger1})
 with $\hat{H} = \hat{T} +
\hat{V}$ as a generalized boson eigenvalue equation,
\begin{equation}
\hat{\cal H}_B \left | \Phi_\lambda \right )
= E_\lambda \hat{ \cal N}_B   \left | \Phi_\lambda \right ).
\label{Schrodinger2}
\end{equation}
Here $\hat{\cal H}_B$ is the boson representation of the fermion Hamiltonian.
Every physical fermion eigenstate in (\ref{Schrodinger1})
has a corresponding eigenstate, with the same eigenvalue, in
(\ref{Schrodinger2}).
Because the space of states constructed from pairs of fermions is
overcomplete, there also exist spurious boson states that do not
correspond to unique physical fermion states.
These spurious states will have zero eigenvalues and so can be identified.

\section{Boson images}

In general the boson representations given in (\ref{CompactNorm}),
(\ref{OneBodyImage}) and (\ref{TwoBodyImage})
do not have good convergence properties,
so that simple termination of the series such as (\ref{NormExpand})
in $\ell$-body
terms is impossible and use of the generalized eigenvalue
equation
(\ref{Schrodinger2}), as written, is problematic.
Instead we ``divide out'' the norm operator to obtain the {\it boson image},
i.e.\ schematically,
\begin{equation}
\hat{h} \sim ``\hat{{\cal H \rm}}_B / \hat{{\cal N \rm}}_B."
\end{equation}
That this is reasonable is suggested by the  explicit forms of
(\ref{OneBodyImage}) and  (\ref{TwoBodyImage}).
The hope of course is that $\hat{h}$ is finite or nearly so, so that a 1+2-body
fermion Hamiltonian is mapped to an image
\begin{equation}
\hat{h} \sim \theta_1 b^\dagger b + \theta_2  b^\dagger b^\dagger b b
+ \theta_3  b^\dagger b^\dagger b^\dagger b b b
+ \theta_4  b^\dagger b^\dagger b^\dagger b^\dagger b b b b + \ldots
\label{HTruncate}
\end{equation}
 with the
 $\ell$-body terms, $\ell > 2$,  zero or greatly suppressed.
We now discuss how to ``divide out'' the norm.

\subsection{Exact results:  Full Space}

It turns out that for a number of cases the image of the
Hamiltonian is exactly finite. In particular, for the full boson Fock space
the representations factor in a
simple way:
$\hat{\cal T}_B = \hat{\cal N}_B \hat{T}_B =  \hat{T}_B \hat{\cal N}_B$ and
$\hat{\cal V}_B = \hat{\cal N}_B \hat{V}_B =  \hat{V}_B \hat{\cal N}_B$,
where the factored operators $\hat{T}_B, \, \hat{V}_B$, which we term  the
boson images of  $\hat{T}, \, \hat{V}$, commute with the norm operator
and have simple forms:
\begin{equation}
 \hat{T}_B =
2\sum_{\sigma \tau}
{\rm tr \,}
\left ({\bf A}_\sigma {\bf T A}_{\tau}^\dagger
 \right)
{b}^\dagger_{\sigma} {b}_\tau,
\end{equation}
\begin{equation}
\hat{V}_B  = \sum_{\mu \nu} \left \langle \mu  \right |
V \left | \nu \right \rangle \left[
{b}^\dagger_\mu {b}_\nu
+ 2 \sum_{\sigma \sigma^\prime} \sum_{\tau \tau^\prime}
{\rm tr \,}
\left (
{\bf A}_\sigma {\bf A}^\dagger_\mu
{\bf A}_{\sigma^\prime} {\bf A}^\dagger_\tau
{\bf A}_\nu {\bf A}^\dagger_{\tau^\prime}
\right )
{b}^\dagger_\sigma {b}^\dagger_{\sigma^\prime}
{b}_\tau {b}_{\tau^\prime}\right]
\end{equation}
The proof of the factorization and commutation requires use of the identities
\begin{eqnarray}
2 \sum_{\alpha} {\rm tr} ({\bf Q A}^{\dagger}_{\alpha}) {\rm tr}
({\bf A}_{\alpha} {\bf R}) =
{\rm tr} ({\bf Q R}) - {\rm tr} ({\bf Q}^T{\bf R}),
\\
2 \sum_{\alpha} {\rm tr}
({\bf Q A}^\dagger_\alpha {\bf R} {\bf  A}_\alpha)
= {\rm tr} ({\bf Q }) {\rm tr} ({\bf R })
 - {\rm tr} ({\bf Q}^T   {\bf R})
\end{eqnarray}
(${\bf Q}^T$ is the transpose of ${\bf Q}$)
which in turn are proved using the completeness relation (\ref{completeness}).
The image Hamiltonian
$ \hat{H}_B = \hat{T}_B + \hat{V}_B$ is the one determined by BZ if one
decomposes the Hamiltonian into multipole-multipole form and then
maps these multipole operators.  As discussed earlier, these BZ multipole
operators are finite in the full space.
This result, and its relation to other mappings, was noted by  Marshalek
\cite{Review,Marshalek}

Thus any boson representation of a Hamiltonian factorizes:
$\hat{\cal H}_B = \hat{\cal N}_B \hat{H}_B$ in the full space.  Since the norm
operator is a
function of the ${\rm SU}(2\Omega)$ Casimir operators it commutes with the
boson
images of fermion operators, and one can simultaneously
diagonalize both $\hat{\cal H}_B$ and $\hat{\cal N}_B$.  Then
Eqn.\ (\ref{Schrodinger2}) becomes
\begin{equation}
 \hat{H}_B\left | \Phi_\lambda \right ) =
E^\prime_\lambda \left | \Phi_\lambda \right ).
\label{Schrodinger3}
\end{equation}
where $E^\prime_\lambda = E_\lambda$
for the physical states, but $E^\prime_\lambda$ for the
spurious states is no longer necessarily zero.
The boson Hamiltonian $\hat{H}_B$ is by construction Hermitian and, if one
starts with at most only two-body interactions between fermions, has at most
two-body
boson interactions.  All physical eigenstates of the original fermion
Hamiltonian will have counterparts
in (\ref{Schrodinger3}).  It should  be clear that transition amplitudes
between physical eigenstates will be preserved. Spurious states will also
exist but, since the norm operator $\hat{\cal N}_B$ commutes with the boson
image
Hamiltonian $\hat{H}_B$, the physical eigenstates and the spurious states
will not admix.  Also the spurious states can be identified because, while
they will no longer have zero energy eigenvalues, they have eigenvalue
zero with respect to the norm operator.

\subsection{Exact Results:  Truncated space}

The boson Schr\"odinger equation (\ref{Schrodinger3}),
though finite, is not of
much use as the boson Fock space is  much larger than the original
fermion Fock space, and we still must truncate the boson Fock space.
Although the representations remain exact under truncation,
the factorization  into the image does not persist in general:
 $\left [ \hat{\cal H}_B \right ]_T
\neq   \left [ \hat{\cal N}_B \right ]_T \left [ \hat{H}_B \right ]_T$.
Again for example  consider the pairing
interaction, even in multiple $j$-shells, and a truncation to just $s$-bosons;
then $[ \hat{H}_B  ]_T = s^{\dagger} s +
{1\over 2 \Omega^2} s^{\dagger} s^{\dagger} s s$ whereas
the appropriate image is actually
$\hat{h} =  s^{\dagger} s -{1\over \Omega} s^{\dagger} s^{\dagger} s s$.
This was recognized by Marshalek \cite{Marshalek}.
(An alternate formulation \cite{Marshalek} does not require the
complete Fock space, but mixes physical and spurious states and so always
requires a projection operator.)

If the  truncated set $\left \{ \bar{\alpha} \right \}$
represents a closed subalgebra, that is,
if the truncated set of fermion pairs are closed under double commutations:
\begin{eqnarray}
\left [ A_{\bar{\alpha}}, A^\dagger_{\bar{\beta}} \right ]
= \delta_{ \bar{\alpha}, \bar{\beta}} - T_{\bar{\alpha} \bar{\beta}}, \\
\left [ T_{\bar{\alpha} \bar{\beta}}, A^\dagger_{\bar{\gamma}} \right ]
= \sum_{\bar{\sigma}}
\Gamma _{ \bar{\alpha} \bar{\beta}; \bar{\gamma}}^{\bar{\sigma}}
A^\dagger_{\bar{\sigma}};
\end{eqnarray}
 then a factorization \cite{Doba}
\begin{equation}
\left [ \hat{\cal H}_B \right ]_T
=  \left [ \hat{\cal N}_B \right ]_T \hat{h}_D
=   \hat{h}_D^\dagger \left [ \hat{\cal N}_B \right ]_T
\label{DysonFactor}
\end{equation}
{\it does} exist, with $\hat{h}_D$ at most two-body, but not
necessarily Hermitian:
\begin{eqnarray}
\left (T_{\bar{\alpha} \bar{\beta}}\right )_D = \sum_{ \bar{\sigma} \bar{\tau}}
\Gamma _{ \bar{\alpha} \bar{\beta}; \bar{\tau}}^{\bar{\sigma}}
b^\dagger_{\bar{\sigma}} b_{\bar{\tau}}, \\
\left ( A^\dagger_{\bar{\mu}}A_{ \bar{\nu}} \right )_D
= b^\dagger_{\bar{\mu}}\left (
b_{\bar{\nu}} - \sum_{\bar{\sigma} \bar{\tau} \bar{\tau}^\prime}
 \Gamma _{ \bar{\nu} \bar{\tau}^\prime; \bar{\tau}}^{\bar{\sigma}}
b^\dagger_{\bar{\sigma}} b_{\bar{\tau}} b_{\bar{\tau}^\prime} \right ).
\end{eqnarray}
We term this a generalized {\em Dyson}
image \cite{Review,RS,Dyson}.  Although
$ \left \langle {\rm physical} \, \left | \hat{h}_D \right | {\rm spurious}
\right \rangle = 0$, if $\hat{h}_D$ is non-Hermitian, that is
$\left [  \left [ \hat{\cal N}_B \right ]_T, \hat{h}_D \right ] \neq 0$,
then unfortunately
$ \left \langle {\rm spurious} \, \left | \hat{h}_D \right | {\rm physical}
\right \rangle  \neq 0$.
 In the full space, of course,
all definitions of boson images coincide and yield the same result.

We have found conditions under which $\hat{h}_D$ is additionally Hermitian
and commutes
with the truncated norm operator $\left [ \hat{\cal N}_B \right ]_T$.
Consider a partition of the single fermion states labeled by $i = (i_a,
i_c)$, where the dimension of each subspace is $2\Omega_a$, $2\Omega_c$  so
that $\Omega = 2 \Omega_a \Omega_c$.  We denote the amplitudes for the
truncated space as ${\bf A}^{\dagger}_{\bar{\alpha}}$ and assume they can be
factored, $ ({\bf A}^{\dagger}_{\bar{\alpha}})_{ij} = ({\bf K}^{\dagger})_{i_a
j_a} \otimes (\bar{{\bf A}}^{\dagger}_{\bar{\alpha}})_{i_c j_c}$, with
${\bf K}^{\dagger} {\bf K} = {\bf K K}^{\dagger} = \frac{1}{2 \Omega_a}$ and
${{\bf K}^T} = (-1)^p {\bf K}$, where $p = 0$ (symmetric) or $p = 1$
(antisymmetric).
Furthermore we assume the completeness relation (\ref{completeness}),
which was crucial for proving
that $\hat{\cal H}_B = \hat{\cal N}_B \hat{H}_B$ \cite{JG2},
is valid for the  truncated space; i.e.,
\begin{equation}
\sum_{\bar{\alpha}} (\bar{\bf A}^{\dagger}_{\bar{\alpha}})_{i_c j_c}
(\bar{\bf A}_{\bar{\alpha}})_{j_c^{\prime} i_c^\prime}
= \frac{1}{2} \left[ \delta_{i_c, i^{\prime}_c} \delta_{j_c ,j^{\prime}_c} -
(-1)^p \delta_{i_c, j^{\prime}_c} \delta_{i^{\prime}_c,j_c} \right].
\label{newcomplete}
\end{equation}
The norm operator in the truncated space then becomes
\begin{equation}
\left[\hat{\cal{N}}_B\right]_T  = \colon {\rm exp} \sum_{k = 2}
\left(\frac{-1}{\Omega_a}\right)^{k-1} \frac{1}{k}
{\rm tr} (\bar{\bf P}^k) \colon ,
\end{equation}
where $\bar{\bf P} = \sum_{\bar{\sigma} \bar{\tau}} b^{\dagger}_{\bar{\sigma}}
b_{\bar{\tau}} \bar{\bf A}_{\bar{\sigma}}
\bar{\bf A}^{\dagger}_{\bar{\tau}}$ so that
 $\left[{ \bf P}\right]_T = \left(\frac{1}{2
\Omega_{a}}\right)\bar{\bf P}$.
In this case the boson image of a one-body operator is the truncation of the
boson image in the full space,
\begin{equation}
\left [ \hat{\cal T}_B \right ]_T =
\left [ \hat{\cal N}_B \right ]_T
\left [ \hat{ T}_B \right ]_T
\end{equation}
\begin{equation}
\left [ \hat{ T}_B \right ]_T =
2\sum_{ \bar{\sigma}, \bar{\tau}}
{\rm tr \,} \left ( {\bf A}_{\bar{\sigma}} {\bf T}
{\bf A}^\dagger_{\bar{\tau}} \right )
b^\dagger_{\bar{\sigma}} b_{\bar{\tau}}.
\end{equation}
The representation of a two-body interaction can be factored into a boson image
 times the truncated norm,
\begin{equation}
\left [ \hat{\cal V}_B \right ]_T =
\left [ \hat{\cal N}_B \right ]_T \hat{v}_D;
\label{DysonImage2}
\end{equation}
however, $\hat{v}_D$, while finite (1+2-body), is not simply related to
$\left [ \hat{V}_B \right ]_T$ as is the case for one-body operators. If one
writes
\begin{equation}
\hat{v}_D =
\sum_{\bar{\sigma}, \bar{\tau}} \langle \bar{\sigma}|V|\bar{\tau} \rangle
b^{\dagger}_{\bar{\sigma}} b_{\bar{\tau}} +
\sum_{\bar{\sigma}\bar{\sigma}^{\prime}\bar{\tau}\bar{\tau}^{\prime}}\langle
\bar{\sigma} \bar{\sigma}^{\prime} |v| \bar{\tau} \bar{\tau}^{\prime} \rangle
b^{\dagger}_{\bar{\sigma}} b^{\dagger}_{\bar{\sigma}^{\prime}} b_{\bar{\tau}}
b_{\bar{\tau}^{\prime}},
\end{equation}
then matrix elements of the two-boson interaction
are
\begin{eqnarray}
\left \langle \bar{\sigma} \bar{\sigma}^{\prime} | v | \bar{\tau}
\bar{\tau}^{\prime} \right \rangle =
\nonumber \\
\sum_{\mu, \nu}\frac{ \left \langle \mu | V | \nu
\right \rangle}{\Omega_a (2 \Omega_a - (-1)^p)(\Omega_a + (-1)^p)} {\rm tr}_a
\lbrace {\rm tr}_c (\bar{\bf A}_{\bar{\sigma}} \bar{{\bf
A}}^\dagger_{\bar{\tau}}
{\bf A}_{\nu} \bar{\bf A}^\dagger_{\bar{\tau}^{\prime}}) {\rm tr}_c
(\bar{\bf A}_{\bar{\sigma}^\prime} {\bf A}^\dagger_{\mu})
\nonumber \\
+ 2 \Omega_{a} [{\rm tr}_c (\bar{\bf A}_{\bar{\sigma}} {\bf A}^\dagger_{\mu}
\bar{\bf A}_{\bar{\sigma}^{\prime}} \bar{\bf A}^\dagger_{\bar{\tau}} {\bf
A}_{\nu} \bar{\bf A}^\dagger_{\bar{\tau}^{\prime}}) - {\rm tr}_c (\bar{\bf
A}_{\bar{\sigma}} \bar{\bf A}^\dagger_{\mu} \bar{\bf A}_{\bar{\sigma}^{\prime}}
 \bar{\bf A}^\dagger_{\bar{\tau}} {\bar{\bf A}}_{\bar{\nu}} {\bf K} \bar{\bf
A}^\dagger_{\bar{\tau}^{\prime}})]
\nonumber
\\
- \Omega_a (2\Omega_a + (-1)^p) {\rm tr}_c ({\bf A}_{\nu} {\bf K}^{\dagger}
\bar{\bf
A}^\dagger_{\bar{\tau}} \bar{\bf A}_{\bar{\sigma}^{\prime}} \bar{\bf
A}^\dagger_{\bar{\tau}^{\prime}}) \delta_{\bar{\sigma}, \mu} \rbrace ,
\label{BigLongEqn}
\end{eqnarray}
where ${\rm tr}_a, {\rm tr}_c$ are traces only in the $a$- and $c$- spaces,
respectively.

Upon inspection one sees the image (\ref{BigLongEqn})
is not constrained to be Hermitian.
Consider the additional condition between the matrix elements of
the interaction:
\begin{eqnarray}
\sum_{\mu, \nu} \left \langle \mu \left | V \right | \nu \right \rangle
\sum_{i_a, j_a}
\left (  {\bf A}_\nu \right )_{ i_a i_c, j_a j_c}
\left ( {\bf A}^\dagger_\mu  \right )_{j_a j_c^\prime, i_a i_c^\prime}
\nonumber \\
= N_a
\sum_{\mu, \nu}
 \left \langle \mu \left | V \right | \nu \right \rangle
\sum_{i_a, j_a}
\left (  {\bf A}_\nu \right )_{ i_a i_c, j_a j_c}
\left ( {\bf K}^\dagger \right ) _{ j_a, i_a}
\sum_{i_a^\prime, j_a^\prime}
\left ( {\bf K} \right ) _{ i_a^\prime, j_a^\prime}
\left ( {\bf A}^\dagger_\mu  \right )_{j_a^\prime j_c^\prime,
i_a^\prime i_c^\prime}
\label{HermCondition}
\end{eqnarray}
where the factor $N_a=\Omega_a(2\Omega_a+(-1)^p) $ is the number of pairs in
the excluded subspace.  While condition (\ref{HermCondition})
looks complicated there are
interactions that satisfy it; for example, two-body interactions constructed
from one-body operators $\hat{V} = \hat{T}_{\bar{\alpha} \bar{\beta}}
\hat{T}_{\bar{\alpha}^\prime \bar{\beta}^\prime}$ where $\hat{T}_{\bar{\alpha}
\bar{\beta}} = \left [A^\dagger_{\bar{\alpha}} , A_{\bar{\beta}} \right ]$.
When (\ref{HermCondition})
is satisfied then $\hat{v}_D$ is Hermitian and although $\hat{v}_D
\neq \left [\hat{V}_B \right ]_T$ they are simply related:
\begin{eqnarray}
\hat{v}_D = \sum_{\bar{\sigma},\bar{\tau}}
\left \langle \bar{\sigma} \left | V \right | \bar{\tau} \right \rangle
b^\dagger_{\bar{\sigma}} b_{\bar{\tau}}
\nonumber \\
 + 2 f_{\Omega_a}
\sum_{{\mu},{\nu}} \left \langle {\mu} \left | V \right | {\nu} \right \rangle
\sum_{  \bar{\sigma} \bar{\sigma}^\prime,
\bar{\tau} \bar{\tau}^\prime }
{\rm tr \,} \left (
{\bf A}_{\bar{\sigma}} {\bf A}^\dagger_{{\mu}}
{\bf A}_{\bar{\sigma}^\prime} {\bf A}^\dagger_{\bar{\tau}}
{\bf A}_{{\nu}} {\bf A}^\dagger_{\bar{\tau}^\prime}
\right )
b^\dagger_{\bar{\sigma}} b^\dagger_{\bar{\sigma}^\prime}
b_{\bar{\tau}} b_{\bar{\tau}^\prime}
\label{HermDysonImage}
\end{eqnarray}
with  $f_{\Omega_a} = 4\Omega_a^2/N_a $ renormalizing the two-boson part of $
\left [ \hat{V}_B \right ]_T$ by a factor  which
ranges from unity (full space) to 2 for a very small subspace.
Not all interactions satisfy (\ref{HermCondition});
for example, the pairing interaction never
does except in the full space.  For the pairing interaction
$\left \langle  {\mu} \left |
V^{\rm pairing} \right | {\nu} \right \rangle = \delta_{\mu, 0}
\delta_{\nu, 0}G$, and
${\bf A}_{0} {\bf A}_0^\dagger = \frac{1}{2 \Omega}$,
and the image (\ref{DysonImage2}) $\hat{v}_{D}^{\rm pairing} $ becomes
(remembering $\Omega = 2 \Omega_a \Omega_c$)
\begin{equation}
 G \left \lbrace \hat{{\rm N}}_0 [ 1 - \frac{2}{\Omega}
\hat{{\rm N}} + \frac{1}{\Omega} + \frac{\hat{{\rm N}}_0}{\Omega}]
 + \sum_{\bar{\tau} \bar{\tau}^\prime \neq 0, \bar{\sigma}} {\rm tr}\;
(\bar{{\bf A}}_{\bar{\sigma}} \bar{{\bf A}}^\dagger_{\bar{\tau}} \bar{{\bf
A}}_0 \bar{{\bf A}}_{\bar{\tau}^\prime}^\dagger) b^\dagger_{\bar{\sigma}}
b^\dagger_0 b_{\bar{\tau}} b_{\bar{\tau}^\prime} \right \rbrace,
\label{PairingImage}
\end{equation}
where $\hat{{\rm N}}$ is the total number of bosons, $\hat{{\rm N}} =
\sum_{\bar{\tau}} b^\dagger_{\bar{\tau}} b _{\bar{\tau}}$, and $\hat{{\rm N}}_0
=  b^\dagger_0 b _0$.  The second term in (\ref{PairingImage})
is not Hermitian but can be
transformed away by a similarity transformation \cite{GT},
leaving the first term as  a finite Hermitian image which gives the correct
eigenvalues for all N.

The SO(8) and Sp(6) models \cite{SO8} belong to a class of models which have a
subspace for which (\ref{newcomplete}) is valid and interactions which
satisfy (\ref{HermCondition}).
In these models the shell model orbitals
have a definite angular momentum $\vec{\bf j}$ and
are partitioned into a pseudo orbital angular momentum $\vec{\bf k}$ and
pseudospin $\vec{\bf i}$,
$\vec{\bf j} = \vec{\bf k}+ \vec{\bf i}$.  The
amplitudes are then given as products of Clebsch-Gordon coefficients,
\begin{equation}
\left(A^{\dagger}_{\alpha}\right)_{ij} = \frac{(1 + (-1)^{K + I})}{2}
(k \, m_i, k\, m_j |K_\alpha \, M_\alpha) \, (i \, \mu_i, i\, \mu_j |I_\alpha
\, \mu_\alpha),
\label{SO8pairDefn}
\end{equation}
 where $K$ and $I$ are the total pseudo orbital angular
momentum and pseudospin respectively of the pair of nucleons.  For the SO(8)
model  ${\bf i} = \frac{3}{2}$ and one  considers the subspace of pairs with $K
 = 0$ $(p = 0)$, $(\bar{A}^{\dagger}_{\bar{\alpha}})_{ij} = \frac{(1 +
(-1)^I)}{2} ( i \, \mu_i, i \, \mu_j | I_\alpha \, \mu_\alpha)$;
in the Sp(6) model  ${\bf k} = 1$ and one considers the subspace with $I = 0$
$(p = 1)$, $(\bar{A}^{\dagger}_{\bar{\alpha}})_{ij} = \frac{(1 + (-1)^K)}{2}
(k \, m_i, k\, m_j | K_\alpha \, M_\alpha)$.
The complicated conditions (\ref{HermCondition})
hold true for important cases, such as the
quadrupole-quadrupole and other multipole-multipole interactions in the
SO(8) and Sp(6) models (that is, interactions of the generic form
$P^r \cdot P^r$ in the notation of \cite{SO8}) where therefore
have Hermitian Dyson images.  Not all interactions in these models
have Hermitian Dyson images.  For example, pairing in any model
(see (\ref{PairingImage})) and, in the SO(8) model, the
combination $V^7 = S^\dagger S + {1\over 4} P^2\cdot P^2 $,
where $S = \sqrt{\Omega} A_{J=0}$, $\Omega =  4k + 2$,
which is the SO(7) limit.
It so happens that these particular cases nonetheless
can be brought into finite, Hermitian form as discussed in the
next section.

\subsection{  Approximate or numerical images \label{ApproxImages} }

The most general image Hamiltonian one can define is
\begin{equation}
\hat{h} \equiv {\cal U} \left [ \tilde{\cal N}_B \right ]_T^{-1/2}
\left [\hat{{\cal H \rm}}_B \right ]_T \left [ \tilde{\cal N}_B \right
]_T^{-1/2} {\cal U}^\dagger,
\label{HImage}
\end{equation}
which is manifestly Hermitian for any truncation scheme and any
interaction, with ${\cal U}$  a unitary operator.
(Because the norm is a singular operator
it cannot be inverted. Instead $\left [ \tilde{\cal N}_B \right ]_T^{-1/2}$
is calculated from the norm only in the physical subspace,
with the zero eigenvalues which annihilate the spurious states retained.
Then $\hat{h}$ does not mix physical and spurious states.)
If $\; {\cal U} = 1$,
this is  the democratic mapping \cite{democratic}. Again, for the
full space $\left[ \hat{\cal N}_B , \hat{H}_B \right] = 0$ and hence
$\hat{h} = \hat{h}_D = \hat{H}_B$.

This prescription is, we argue, useful for a practical derivation of boson
image Hamiltonians.  Ignoring for the moment the unitary transformation
${\cal U}$, consider the expansion (\ref{HTruncate}) of $\hat{h}$.
The operators $\left [\hat{{\cal H \rm}}_B \right ]_T$ and
$\left [ \tilde{\cal N}_B \right ]_T^{-1/2}$ have similar expansions, and by
multiplying out (\ref{HImage})
one sees immediately that the coefficient $\theta_\ell$ depends only
on up to $\ell$-body terms in
$\left [ {\cal H}_B \right ]_T$ and
$\left [ \tilde{\cal N}_B \right ]_T^{-1/2}$, derived from $2\ell$-fermion
matrix elements which are tractable for $\ell$ small.
Ideally $\hat{h}$ would have at most two-body terms,  and
our success in finding finite images in the previous section gives us
hope that the high-order many-body terms may be small; at any
rate the convergence can be calculated and checked term-by-term.
Specifically, consider the convergence
of the series (\ref{HTruncate}) as a function of $\ell$. A rough estimate
is that, for an $N$-boson Fock space, one can truncate to the $\ell$-body
terms if for $\ell^\prime > \ell$,  $\theta_{\ell^\prime}$
is sufficiently small compared to
$\theta_\ell \times (N-\ell^\prime)!/(N-\ell)!$; the strictest condition is to
require $\theta_{\ell^\prime} \ll \theta_\ell /(\ell^\prime-\ell)!$.

Although we have given in section IV analytic expressions for the coefficients
of the boson representations, in practice one only needs the fermion matrix
elements, given in section III, for the norm and the Hamiltonian or other
operators.  The coefficients of the image $\hat{h}$
are then found by numerical induction. Suppose one has the coefficients
of $\hat{h}$ up to
$\ell$-boson terms.  One takes the matrix elements of
$\left [ \tilde{\cal N}_B \right ]_T^{-1/2}$---note
that in calculating $\left [ \tilde{\cal N}_B \right ]_T^{-1/2}$, one
{\em first} truncates and {\em then} calculates the inverse-square-root; the
two operations do not commmute!---and of the Hamiltonian or transition
operator in the  $2(\ell+1)$ fermion space, and multiply out those matrices
as in (\ref{HImage}), yielding the matrix elements of $\hat{h}$ in
the $\ell+1$-boson space.  The coefficients $\theta_{\ell+1}$ of $\hat{h}$ in
(\ref{HTruncate}) are then uniquely determined, up to the freedom embodied in
the unitary transform ${\cal U}$.

The Hermitian image $\hat{h}$, defined in (\ref{HImage}),
is related to the Dyson image  $\hat{h}_D$, defined in (\ref{DysonFactor}),
by a similarity transformation
${\cal S} = {\cal U} \left [\tilde{\cal N}_B \right ]_T^{1/2}$,
\begin{equation}
\hat{h} = {\cal S} \hat{h}_D {\cal S}^{-1}.
\end{equation}
 The similarity transformation
${\cal S}$ orthogonalizes the fermion states
$\left | \Psi_{\bar{\alpha}} \right \rangle$  inasmuch
$({\cal S }^{-1})^\dagger\tilde{\cal N }_B{\cal S}^{-1} = 1$
in the physical space (and $=0$ in the spurious space). This
is akin to Gram-Schmidt orthogonalization and the freedom to choose ${\cal U}$,
and ${\cal S}$, corresponds to the freedom one has in ordering the
vectors in the Gram-Schmidt procedure.  The OAI and democratic mappings
are just two particular choices out of many; the former orders the states by
seniority whereas the latter takes ${\cal U}=1$.

We can use this freedom in the choice of ${\cal U}$ to our advantage,
which we illustrate in the SO(8) model \cite{SO8}, truncating to the
space of $K=0$ pairs (see eqn.~(\ref{SO8pairDefn})), for which one has
only an $I = J = 0$ pair (mapped to an $s$ boson) and a quintuplet of $I=J=2$
pairs (mapped to  $d_m$ bosons, $m = -2 \ldots 2$).  To second order,
the norm is
\begin{eqnarray}
\hat{\cal N}_B = 1 -{ 1\over 2\Omega} s^\dagger s^\dagger ss
- {1 \over 2\Omega} \sum_{L = 0,2,4,} (2 - 5\delta_{L,0} )
\left [ d^\dagger \otimes d^\dagger \right ]_L \cdot
\left [ \tilde{d} \otimes \tilde{d} \right ]_L
\nonumber \\
-{ \sqrt{5} \over 2 \Omega}
\left \{ s^\dagger s^\dagger \left [ \tilde{d} \otimes \tilde{d} \right ]_0
+ \left [ d^\dagger \otimes d^\dagger \right ]_0 s s \right \}
-{2 \over \Omega} \hat{n}_d \hat{n}_s.
\end{eqnarray}
We pay particular attention to three interactions which correspond
to algebraic limits: the pure pairing interaction
$V^{\rm pairing} = S^\dagger S = \Omega A^\dagger_0 A_0 $, the
quadrupole-quadrupole interaction
$V^{QQ}$ (= $P^2 \cdot P^2$ in the notation of \cite{SO8}),
which can be written in terms of SO(6) Casimir operators,
 and the linear combination of pairing and quadrupole
$V^7 = V^{\rm pairing} + {1 \over 4}V^{QQ}$
which can be written in terms of SO(7) Casimirs.
As discussed in the previous section, the Dyson image of $V^{QQ}$
is Hermitian and finite, and hence $\hat{h}_D = \hat{h}$ with
${\cal U} = 1$:
\begin{equation}
\left ( V^{QQ} \right )_D  = 20 s^\dagger s + 4 d^\dagger \cdot \tilde{d}
+ 4 \sqrt{5}
\left \{ s^\dagger s^\dagger \left [ \tilde{d} \otimes \tilde{d} \right ]_0
+ \left [ d^\dagger \otimes d^\dagger \right ]_0 s s \right \}
+ 8 \hat{n}_d \hat{n}_s.
\label{QQimage}
\end{equation}
 The Dyson images of
the pairing and SO(7) interactions are finite but non-Hermitian:
\begin{eqnarray}
\left ( V^{\rm pairing} \right )_D  =
\Omega s^\dagger s - s^\dagger s^\dagger s s
- \sqrt{5}
 s^\dagger s^\dagger \left [ \tilde{d} \otimes \tilde{d} \right ]_0
-2 \hat{n}_d \hat{n}_s,
\\
\left ( V^{7} \right )_D  =
(\Omega+5) s^\dagger s + d^\dagger \cdot \tilde{d}
- s^\dagger s^\dagger s s
+ \sqrt{5}  \left [ d^\dagger \otimes d^\dagger \right ]_0 s s.
\end{eqnarray}
We arrived at these images by computing the fermion matrix elements
using the methods of \cite{SO8} and writing the norm and the representations
of the interactions as matrices, which in second order are at most
$2 \times 2$, and then directly multiplied the
matrices ${\cal N}^{-1} {\cal H}$.
We went to third order to confirm the Dyson images are finite.
Using these same matrices we could also calculate
$ {\cal N}^{-1/2} {\cal H} {\cal N}^{-1/2}$, taking ${\cal U} = 1$,
which yields again (\ref{QQimage}) for $V^{QQ}$;
for $V^{\rm pairing}$ and $V^7$ the images are then Hermitian but
with nonzero third-order, and presumably higher-order, terms.

We then found ${\cal U}$'s $\neq 1$ for both the pairing and SO(7) cases
(but not the same ${\cal U}$) such that
their respective Hermitian images $\hat{h}$ are finite; the one for pairing
is unsurprisingly the OAI prescription, while that for SO(7) is the opposite,
orthogonalizing states of low seniority against states of higher
seniority.  These finite, Hermitian images are:
\begin{eqnarray}
v^{\rm pairing}  =
\Omega s^\dagger s - s^\dagger s^\dagger s s ,
\\
 v^{7}  =
(\Omega+5) s^\dagger s + d^\dagger \cdot \tilde{d}
- s^\dagger s^\dagger s s .
\label{V7exact}
\end{eqnarray}
Note that these general Hermitian images are not simply related
to the truncation of the full image through an overall
renormalization, as is the Hermitian Dyson image
(\ref{HermDysonImage}).  For the pairing interaction, the
coefficient of the $s^\dagger s^\dagger s s$ term
changes sign, while off-diagonal
terms such as $d^{\dagger} d^{\dagger} ss$, which exist in the image
in the full space, vanish, and in some cases like SO(8) other diagonal
terms such as $d^{\dagger} d^{\dagger} \tilde{d} \tilde{d}$ can also vanish
(depending on the model space).

If one uses an ``inappropriate'' transform ${\cal S \rm}$ it can induce
an unneeded and unwanted many-body dependence.  This principle we
illustrate  in  the SO(8) model, by deriving up to second order
the Hermitian image of $V^7$ but using ${\cal U}_{OAI}$:
\begin{eqnarray}
v^7_{OAI} = (\Omega+5) s^\dagger s + d^\dagger \cdot \tilde{d}
- \left (1 + { 10 / \Omega \over 1 - {1 \over \Omega} } \right )
 s^\dagger s^\dagger s s
- { 3/ \Omega \over  1- {1\over \Omega} }
\left [ d^\dagger \otimes d^\dagger \right ]_0 \cdot
\left [ \tilde{d} \otimes \tilde{d} \right ]_0
\nonumber \\
- { 2 \sqrt{5} \over 1 - {1 \over \Omega} }
\sqrt{ \left (1 -{2 \over \Omega} \right )\left( 1 + {4 \over \Omega} \right)}
 \left \{ s^\dagger s^\dagger \left [ \tilde{d} \otimes \tilde{d} \right ]_0
+ \left [ d^\dagger \otimes d^\dagger \right ]_0 s s \right \}
\label{V7OAI}
\end{eqnarray}
plus higher order terms which we drop; this is equivalent to the
standard OAI procedure \cite{OAI} computed in the 2-boson space.
In figure 1 we
display the spectra of (\ref{V7exact}) on the left, which is
the exact SO(7) result, and (\ref{V7OAI}) on
the right, taking $\Omega = 10$ and $N=7$.
The distortions in the right-hand spectrum from the exact result, such as
the overall energy shift and the large perturbation in the third band,
indicates the importance of the missing many-body terms.  In OAI \cite{OAI}
these many-body terms would appear implicitly in the
$N$-dependence of the coefficients for the two-body terms.
{}From the existence of (\ref{V7exact}) we see that this strong OAI
$N$-dependence is, for this case at least, a needless complication.
Therefore it is possible that some of the $N$-dependence of OAI is induced by
their choice of orthogonalization and could be minimized with a
different choice.  We are currently exploring how to exploit this freedom to
best effect.

\section{Effective Representations and Images}

So far we have considered the mapping of a Hamiltonian in a truncated fermion
space to a boson space.   That truncated fermion space, however, may be
inadequate for reproducing the spectrum of the full space.  Consider
a single $j$-shell fermion space. If the interaction is dominated by
pairing, then truncation to just $J=0,2$ ($s,d$) pairs is reasonable\cite{OAI};
but for other interactions, particularly the quadrupole-quadrupole
interaction, the $s,d$-space is inadequate\cite{QQ17}. To rectify these
shortcomings one must introduce an effective interaction theory for
boson mappings.

An important issue is at what stage to introduce effective operators.
For example, one could start from an effective fermion Hamiltonian.
Because the fermion-pair basis is non-orthogonal, however, computation of
the effective fermion interaction would be tricky; furthermore, the
starting Hamiltonian would have a number dependence which would be
difficult to separate from the number dependence induced by the subsequent
boson mapping.  At the other extreme, Sakamoto and Kishimoto \cite{Sakamoto}
start from a boson image and use perturbation theory to account for
excluded states.   They do not rigorously derive their effective interaction
and it is not clear that they properly account for the exchange terms, etc.,
included in the norm operator.

A better, intermediate approach, is to
calculate the boson image in a larger space---say $sdg$---and then renormalize
the pure $sd$ interaction so as to account for the effect of the
$g$-boson \cite{grenorm}. Ideally, one should start with a sufficient number
of species of bosons so as to exactly span the fermion Fock space, and
then further truncate the space and renormalize.
The number of species required will depend on the number of
bosons (or pairs) $N$, however, and we know of no prescription for determining
this set of bosons (except for $N=1$ when all are necessary).

We present a rigorous and general approach to effective interactions. Following
the usual Feshbach \cite{RS} derivation, we partition the
{\it boson}
Fock space using $P$ to project out the allowed space and $Q$ its
compliment, with $P+Q=1$, $P^2 =P$, $Q^2=Q$ and $PQ =QP = 0$.
Then the truncated representations are simply
\begin{eqnarray}
\left [ {\cal H}_B \right ]_T = P {\cal H}_B P, \\
\left [ {\cal N}_B \right ]_T = P {\cal N}_B P,
\end{eqnarray}
and $\left | \Phi \right )_T = P \left | \Phi \right )$.  Then
the generalized eigenvalue equation in the full space
becomes
\begin{equation}
\left[ \hat{\cal H}_B \right ]_T^{\rm eff}(E_\lambda)
\left | \Phi_\lambda \right )_T
= E_\lambda \left [\hat{ \cal N}_B
\right]_T^{\rm eff}(E_\lambda)  \left | \Phi_\lambda \right )_T,
\nonumber
\end{equation}
with
\begin{eqnarray}
\left[ \hat{\cal H}_B \right ]_T^{\rm eff}(E)
=P \hat{\cal H}_B P + P
\hat{\cal H}_B Q {1 \over Q(  E\hat{\cal N}_B-  \hat{\cal H}_B  )Q }
Q \hat{\cal H}_B P  \nonumber \\
- P \hat{\cal H}_B Q {E \over Q(
E\hat{\cal N}_B -  \hat{\cal H}_B )Q } Q \hat{\cal N }_B P
- P \hat{\cal N}_B Q {E \over Q(  E\hat{\cal N}_B - \hat{\cal H}_B )Q } Q
\hat{\cal H }_B P
\nonumber \\
+P \hat{\cal N}_B Q {E^2 \over Q(  E
\hat{\cal N}_B - \hat{\cal H}_B  )Q } Q \hat{\cal N}_B P
+ E{\cal A}(E),
\label{Effective}
\\
\nonumber \\
\left[ \hat{\cal N}_B \right ]_T^{\rm eff}(E)
=P \hat{\cal N}_B P + {\cal A}(E). \nonumber
\end{eqnarray}
One can also in principle construct energy-independent, but non-Hermitian,
effective representations \cite{EffBoson}.
There is some ambiguity in the definition of the effective representations as
denoted by ${\cal A}(E)$. For example, one could define
$\left[ \hat{\cal H}_B \right ]_T^{\rm eff}$ to be simply
$P \hat{\cal H}_B P$ with the remaining terms in (\ref{Effective}) absorbed
into the definition of $\left[ \hat{\cal N}_B \right ]_T^{\rm eff}(E)$, and
in principle ${\cal A}(E)$ could be anything at all.

Now consider boson images, where one divides out the norm operator.
We suggest that effective operator theory may be more efficient
when applied to representations rather than images, by which we mean that
the corrections are smaller.  Suppose one started with the image Hamiltonian
in the full space, $H_B$ as defined previously, and from that constructed
an effective image in the usual way,
\begin{equation}
H_B^{\rm eff} =
PH_B P + P H_B Q { 1 \over Q( E-H_B)Q } Q H_B P
= \left [ H_B \right ]_T + \Delta H_B(E).
\end{equation}
Now compare that with the effective image constructed from effective
representations,
\begin{equation}
h_B^{\rm eff}(E) \equiv
\left ( \left [ \tilde{\cal N}_B \right ]_T^{\rm eff}(E) \right) ^{-1/2}
\left [ {\cal H}_B \right ]_T^{\rm eff}
\left ( \left [ \tilde{\cal N}_B \right ]_T^{\rm eff} (E)\right) ^{-1/2}
= \hat{h}_B + \Delta h_B(E)
\end{equation}
(leaving aside the issue of the choice of an overall unitary
transformation ${\cal U}$). Now
$\hat{h}_B \neq \left [ H_B \right ]_T$.  Which approach is better?  In
those cases, such as SO(8) and
Sp(6), where, the $P$-space decouples completely from the $Q$ space,
$\Delta h_B = 0$ but $\Delta H_B$ cannot be zero.
Hence the corrections $\Delta h_B$ from using
effective representations  can be smaller than the corrections $\Delta H_B$
determined from performing effective operator theory directly on the image.

We now would like to speculate on the possible use of the
ambiguity operator ${\cal A}(E)$.
In effective operator theory the eigenstates are no longer orthogonal because
of the truncation of the model space; this is expressed by the fact that
one uses either energy-dependent or non-Hermitian effective interactions.
We propose that this non-orthogonality could also be embedded in the choice
of ${\cal A}(E)$, so that the similarity transform on the basis is now
${\cal U}\sqrt{\left [ \hat{\cal N}_B \right ]_T +
{\cal A}(E)}\left | \Phi_{\bar{\alpha}} \right )$.
The ambiguity operator ${\cal A}(E)$ could be chosen so as to minimize
the energy dependence of the final boson image.  Although the similarity
transformation is now energy-dependent, this would not show up in the
calculation of the spectrum, but only in the calculation of effective
transition operators.  These speculations need to be explored in greater
detail.

\section{ Summary}

In order to investigate rigorous foundations for the phenomenological
Interacting Boson Model,
we have presented a rigorous microscopic mapping of fermion pairs to bosons,
paying special attention to exact mapping of matrix elements, Hermiticity,
truncation of the model space, and many-body terms.
First we presented new, general and compact forms for boson
{\it representations} that preserve fermion matrix elements. We then
considered the
boson {\it image} Hamiltonian which results from ``dividing out'' the
norm from the representation; in the full boson Fock
space the image is always finite and Hermitian; in addition we discussed
several analytic cases for truncated spaces where the image is also
finite and Hermitian.  Next, we gave a prescription
which is a generalization of both the OAI and democratic mappings;
in the most general case for  truncated spaces the Hermitian image
Hamiltonian may not be finite but we have demonstrated there is some
freedom in the mapping that one could possibly exploit to minimize the
many-body terms. This freedom, which manifests itself in a similarity
transformation that orders the orthogonalization of the underlying fermion
basis, depends on the Hamiltonian.  Finally, we discussed effective operator
theory for boson mappings.

This research was  supported by the U.S.\ Department of Energy.
The boson calculations for Figure 1 were performed using the PHINT package of
Scholten \cite{Scholten}.

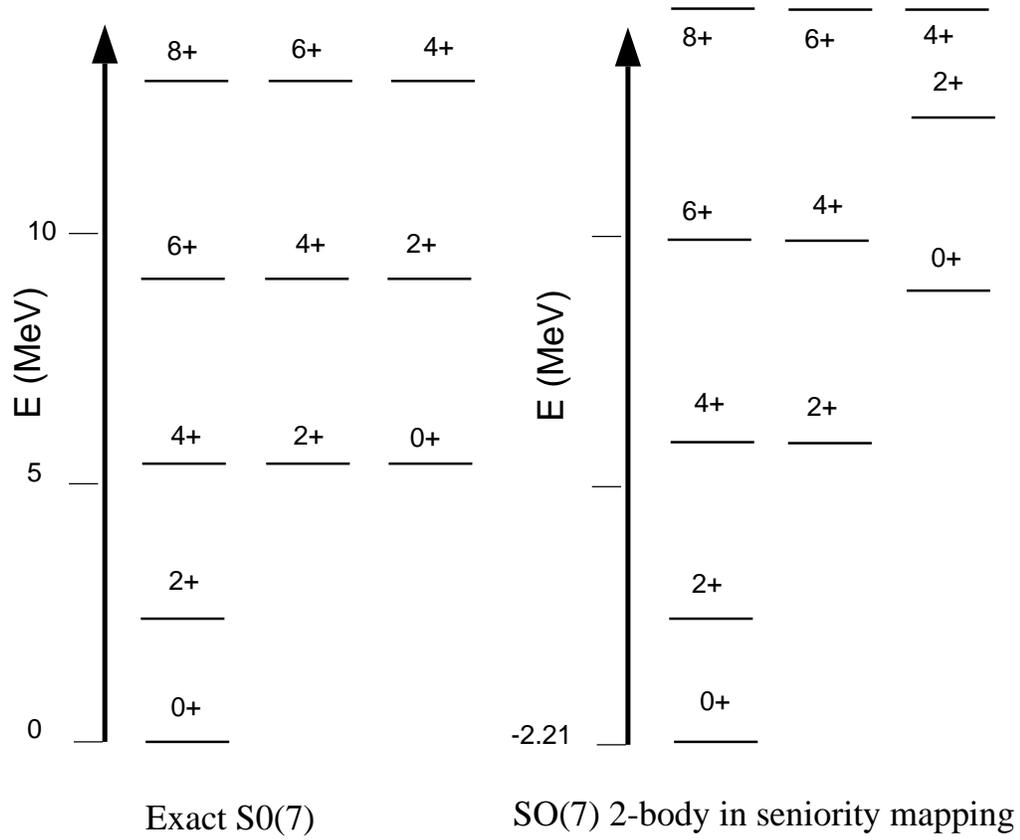
\begin{figure}
\caption{ Spectrum of SO(7) interaction, for 7 bosons,
in SO(8) model with exact
(left) and approximate (right) two-body boson Hamiltonians.}
\label{fig1}
\end{figure}

\end{document}